\documentclass[twocolumn]{aastex62}
\usepackage[utf8]{inputenc}
\usepackage{natbib, xspace, xcolor}
\usepackage{amsmath, paralist}
\usepackage{multirow}
\usepackage[caption=false]{subfig}
\date{\today}

\shorttitle{Second Generation Stars}
\shortauthors{Wells and Norman}

\newcommand{\msun}{{\rm M}_\odot\xspace}

\newcommand{\piii}{Population III\xspace}
\newcommand{\pii}{Population II\xspace}

\newcommand{\gii}{Population II.1\xspace}

\newcommand{\systems}{PIII associations\xspace}
\newcommand{\system}{PIII association\xspace}
\newcommand{\ngii}{4,546\xspace}
\newcommand{\npiii}{3,313\xspace}

\begin{document}

\title{Connecting Primordial Star Forming Regions and Second Generation Star Formation in the Phoenix Simulations}

\author{Azton I. Wells}
\email{aiwells@ucsd.edu}
\affil{Center for Astrophysics and Space Sciences\\
University of California, San Diego, La Jolla, CA, 92093}
\author{Michael L. Norman}
% \author{Britton Smith}
\affil{Center for Astrophysics and Space Sciences\\
University of California, San Diego, La Jolla, CA, 92093}
\affil{San Diego Supercomputer Center\\
University of California, San Diego, La Jolla, CA, 92093}
\begin{abstract}
We introduce the {\em Phoenix Simulations}, a suite of highly resolved cosmological simulations featuring hydrodynamics, primordial gas chemistry, Population III and II star formation and feedback, UV radiative transfer, and saved outputs with $\Delta t$=200 kyr. The suite samples 73,523 distinct primordial star formation events within \npiii distinct regions, forming \ngii second-generation enriched star clusters by $z \geq 12$ within a cumulative 156.25 Mpc$^3$ volume.  The regions that lead to enriched star formation contain up to $167$ primordial stars, with 78.7 \% of regions having experienced multiple types of primordial supernovae.  The extent of a primordial region, measured by its metal-rich surrounding cloud, is highly variable: the average region has radius $\sim 3$ kpc, with 95 \% confidence limit on the distribution of measured radii is $\sim 5-7$ kpc. For continuing star formation, we find that the metallicity distribution of second generation stars is similar to that of subsequent Population II star formation, with both distributions spanning hyper metal-deficient ([Z/H]$\sim-7$) to super-solar ([Z/H]$\sim0.8$). We find that the metallicity of second generation stars has no strong dependence on the configuration of progenitor supernovae, with the mean metallicity of second-generation stars having $-1.73 < $[Z/H]$<-2.15$. Finally, we create an interpretable regression model to predict the radius of metal-rich influence of \piii star systems within the first 7-18 Myr after the first light. The model predicts the radius with $R_2 \gtrsim 0.4$ and mean squared error $\leq 0.06$. The probability distribution function of predicted radii compares well to that of observed radii with Jensen-Shannon distance $\lesssim 0.2$ for all modelled times.  
\end{abstract}

\section{Introduction}
    \label{sec:introduction}
    The first galaxies form from gas that has been enriched by an earlier generation of \piii supernovae. The initial conditions of metallicity in the universe after these first supernova events remains a difficult problem to model in astrophysical simulations.  Researchers conducting astrophysical simulations have three practical options to determine the initial metallicity field prior to enriched (Population II) star formation: assume a metallicity floor (e.g., \citealt{Hopkins2018}); assume initial star formation rates that are independent of metallicity (e.g., \citealt{vogels2013,vogels2014}); or explicitly simulate the primordial (\piii) star formation and feedback \citep{smith2015, xu2016,wise2012,wise2012b}.  Ideally, all researchers would choose the last option, however the extreme small scale ($\sim$ pc$^3$) of primordial molecular cloud formation \citep{ABN00,bromm2002} is at odds with the $\sim$ Gpc$^3$ scale necessary to gain useful statistics of the observable universe; any simulation using current computing facilities that can fully resolve \piii star formation is severely limited in volume, with the largest being only $\sim 300$ Mpc$^3$ \citep{xu2016}.

    The Phoenix (PHX) suite of simulations is designed to facilitate the exploration of a fourth option to model \piii star formation and feedback: to develop surrogate models based on deep neural networks (DNN) as a new sub-grid method to create an heterogeneous metallicity initial condition that reflects the spatially irregular formation of \piii star formation and feedback.  Data from the PHX suite have already been used to train a predictive DNN based surrogate model ({\tt StarFind}) to identify \piii star formation sites without resorting to halo-finding or pc-scale resolution ~\citep{wells2021}, and feature time resolution such that any star formation or feedback event is recorded to disk with 200 kyr time resolution. 
    
    Although their extreme time resolution was intended to provide high resolution of star formation and feedback events for training DNNs, the PHX suite also provides a unique opportunity to study the transition between \piii and second generation (\gii)\footnote{\gii is our designation for the first generation of metal-enriched star formation that occurs in gas enriched exclusively by \piii supernovae.} star formation. Studying low metallicity stars or damped Lyman-alpha systems in observations is currently our only window into the \piii initial mass function (IMF) \citep{cooke2017, welsh2019, welsh2020}, but the uncertainty in the IMF \citep{nakamura2002, ishigaki2018} and metallicity of \gii stars that could form from enriching events make inferences about the \piii era difficult. Due to the fine time resolution in the PHX suite outputs, we use them to study the formation state of small \pii star clusters, as well as the evolution of \piii star forming regions.  This will guide future studies that connect \piii and \pii stars by providing a reference of how many \piii stars can be related to a \pii cluster, as well as the range of metallicities that may be observable in a second generation star.

    The remainder of this paper is organized as follows: Section~\ref{sec:simulations} presents a summary of the simulations and included physical models; Section~\ref{sec:general_results} showcases summary statistics such as star formation rates and halo mass functions of the final simulation states; Section~\ref{sec:analysis} presents a time-resolved study to determine the origins of the first generation of enriched star formation as well as quantification of primordial stellar systems; Section~\ref{sec:model} presents an interpretable regression model that predicts the region of influence for a \piii system, given the stellar masses and birth times within that system; Finally, Section~\ref{sec:conclusion} consolidates our results and presents final notes on both this study and future directions.
    
\section{The Phoenix Simulations}
    
    \label{sec:simulations}
    \begin{figure*}
          
        \includegraphics[width=\textwidth]{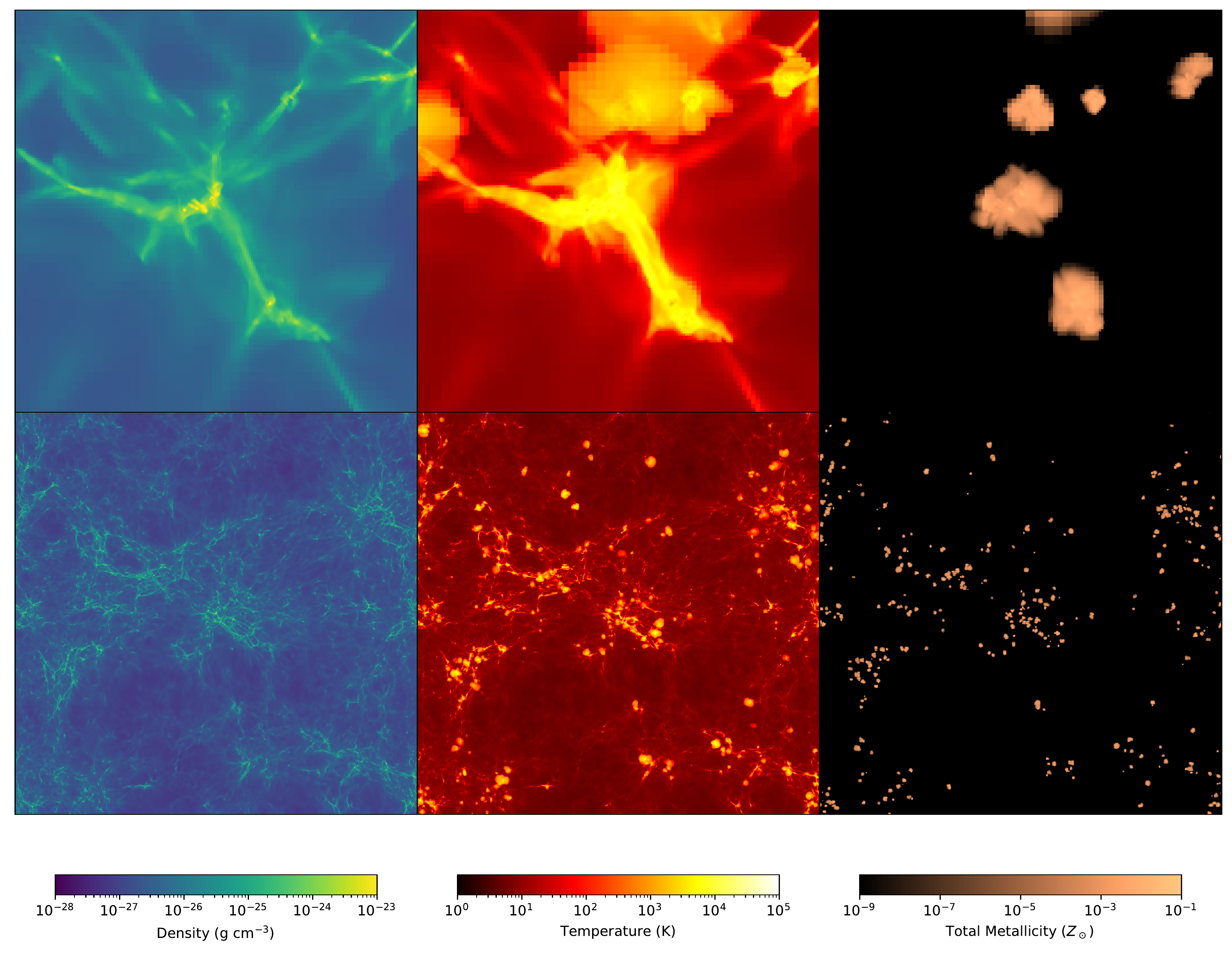}
        \caption{Projections of PHX512 simulation at $z=14.04$. Top row: Projections through a 15 comoving kpc volume surrounding a complex of star forming halos.  Bottom row: 1 Mpc thick slab projection of the entire simulation domain ($L_{box}=5.12$ Mpc) at the same redshift centered on the star forming halo in the top row.}
        \label{fig:sim_prj}
    \end{figure*}

    The PHX suite consists of three simulations performed with the ENZO \citep{enzo} adaptive mesh refinement hydrodynamic cosmology code designed to study the time and spatial evolution of early structure formation and galaxy assembly. Projections of the largest ($512^3$ root-grid) simulation  are shown in Figure \ref{fig:sim_prj}. Prior efforts have produced simulations with similar redshifts as the target range of the PHX suite \citep{wise2012, wise2012b, xu2016}, however the saved outputs of these earlier works have time resolution of $\gtrsim 1-5$ Myr, whereas the PHX suite has 200 kyr between all outputs for redshifts $30 < z < 10$.  
    
    All simulations share identical cosmological parameters with $\{\Omega_m = 0.3111, \Omega_b = 0.048975, \Omega_k = 0, \Omega_\lambda = 0.6889, H_0 = 0.6766, \sigma_8 = 0.811, n = 0.965 \}$~\citep{Planck2014}.  The cosmological initial conditions are generated at $z=99$ using {\tt MUSIC}~\citep{hahn2011}, where each simulation uses a unique random seed to generate an initial state that is consistent with the given cosmological parameters.  All simulations have identical mass and spatial resolution and the same refinement criteria as the {\em Renaissance Simulations} \citep{xu2013}, with dark matter particle mass $M_{DM} = 2.34 \times 10^4 ~\msun$, initial average baryon mass per cell $M_{b,i} = 1.17\times 10^3 ~\msun$.  The root grid can be refined up to 9 levels of adaptive mesh refinement (AMR), where refinement occurs on dark matter density, baryon density, and regions surrounding \piii star particles such that the supernova radius (10 pc) is resolved by at least 4 cells, or is at the maximum AMR level.  Refinement on densities is super-Lagrangian, where cells are flagged for refinement at level $l$ where the cell mass $M_{c} \geq M_{i} \times 2^{-0.4l}$, where $M_i$ refers to $M_{DM}$ or $M_{b,i}$ for refinement on dark matter or baryon density respectively. With these resolution parameters, and assuming $\sim 100$ dark matter particles for a resolved dark matter halo, the least massive resolved halos have virial masses $2.34\times 10^6$ $\msun$, while the most massive halos ($\approx 10^9 \msun$) is limited by the total mass within the volume: $3.93 \times 10^{11}$ M$_\odot$ and $3.14\times 10^{12}$ M$_\odot$ for 256$^3$ and $512^3$ root grids respectively.  The only difference between the three simulations is the box size, and hence the root-grid dimension. The smaller (PHX256-x) have 256$^3$ root-grid cells with spatial dimension (2.56 Mpc)$^3$ while the PHX512 has 512$^3$ root-grid cells spanning (5.12 Mpc)$^3$. The finest spatial resolution achieved on level 9 subgrids is 19.53 comoving pc, which yields $\lesssim$ 2 pc resolution at $z > 8.765$. 
    
    For hydrodynamic and chemical evolution, each simulation includes 9 species non-equilibrium chemistry for primordial gas species H, H$^+$, H$^-$, H$_2$, H$_2^+$, H$_2^{++}$, He, He$^+$, He$^{++}$, and e$^-$ with radiative heating, cooling, and metal-line cooling as in \cite{smith2008}, with hydrodynamics evolved using the piecewise parabolic method~\citep{colella1984}.  Radiation interactions are included via a uniform, redshift-dependent Lyman-Werner H$_2$ dissociating radiation background as documented in \cite{xu2016} as well as photo-dissociating and ionizing radiation from point-sources using the rates and MORAY ray-tracing solver method in \cite{wise2011}.  The radiation is coupled to chemistry through heating and ionization rates, and to hydrodynamic evolution via momentum coupling from photons to the gas.
    
    Each simulation includes two types of star formation events: individual \piii stars and \pii star clusters, each tracked with a star particle. Each type of star particle contributes to a unique metal density field so that the contributions from \piii and \pii stars can be tracked independently.  The star formation and feedback algorithms used in the Phoenix suite are well documented\footnote{https://enzo.readthedocs.io/en/latest/} and described in several prior works~\citep{wise2012, wise2012b, hicks2021}, so the following is restricted to a high-level overview that includes the specific parameters used in the PHX suite.  At each time step, every grid cell is evaluated for star formation according to the following criteria: 
    \begin{inparaenum}[1)]
        \item baryon number density $n_b > 100$,
        \item for \piii stars, H$_2$ fraction $n_{H2}/n_H > 10^{-4}$,
        \item for \piii stars, metallicity\footnote{with metal mass $M_z$ and cell baryon mass $M_c$, metallicity is given by $Z = \log(M_z/M_c) - \log (M_{z,\odot}/\msun)$} $Z < Z_{c}$ for $Z_c = -5.5$,
        \item AMR Grid level is most refined for that location,
        \item the freefall time should be less than the cooling time, and
        \item converging gas flow ($\nabla \cdot \vec\nu < 0$).
    \end{inparaenum}
    If a cell qualifies for \piii star formation, a particle representing a single star is formed centered on the host cell with mass taken from a modified Salpeter initial mass function (IMF)
    \begin{equation}
            f(\log M)dM = M^{-1.3}\text{exp}\bigg[-\bigg(\frac{M_{\mathrm{char}}}{M}\bigg)^{1.6}\bigg]dM,    
            \label{eqn:mass_fn}
    \end{equation} 
    with characteristic mass $M_{char} = 20$ M$_\odot$ with masses in the range $1 \leq M_*/\msun \leq 300$.  The final mass contributing to the star formation is taken from the grid in a sphere containing twice the mass of the star.  If, however, the cell meets all star-forming criteria, but has $Z> Z_c$, a particle representing a \pii star cluster is formed.  The mass of the cluster is drawn from the surrounding cold gas mass, $M_{\rm cold}$, estimated as $7\%$ of the gas mass in a sphere with mean gas number density $n_b > 10^3$ cm$^{-3}$.  The metallicity of the formed star is taken as the mass-averaged metallicity of the cells that contributed to its formation, therefore the final metallicity may be lower or higher than the cell that initially qualified for cluster formation.  The minimum mass of \pii clusters is set to 1000 $\msun$, however, if the mass is not met within a dynamical time after star formation, a low-mass particle is formed with $M_* = 0.07 M_{\rm cold}$ to prevent loss of the ionizing radiation due to lower-mass star clusters.
    
    Stellar feedback for \piii stars is included in two forms: supernovae of varying mass, and point-source radiative feedback.  The supernova channel includes Type-II supernovae (SNe, $11 < M_*/\msun < 20$), hypernovae (HNe, $20\leq M_*/\msun < 40$), and pair-instability supernovae (PISNe, $140 < M_*/\msun < 260$).  For SNe, the ejecta mass, energy, and metal yields are taken from \cite{nomoto2006}; HNe event energy and metal yields are linearly interpolated from these values.  PISNe have ejecta mass, metal yield, and energy taken from \cite{heger2002}.  For each type of supernova, the resulting mass, metal yield, and energy are deposited to the grid in a sphere of 10 pc, or a cube of $3^3$ cell-widths if 10 pc is unresolved.  
    
    \pii stars, modeled as coeval clusters of stars, use a continuous injection model of energy, mass, and metal deposition that represents both supernova and stellar winds.  At each time step during the 20 Myr lifetime of the cluster, mass is returned to the computational grid as 
    \begin{equation}
        m_{\rm ej} = \frac{0.25\Delta t\times M_*/\msun}{t_0 - 4~{\rm Myr}},
    \end{equation}
    where $t_0 \leq 20$ Myr (the lifetime of the particle), and the ejecta has a metallicity fraction matching solar metallicity ($Z_\odot=0.01295$).  The ejecta has energy $1.12\times 10^{49}$ erg/M$_\odot$, which is coupled to the grid as thermal energy along with mass and metal ejecta in a 10 pc sphere surrounding the source particle, again depositing to a $3^3$ cube if if 10 pc is unresolved.

\section{General Observations from the Phoenix Suite}
    \label{sec:general_results}
    % \subsection{Global simulations statistics}
%%%%%%%%%%%%%%%%%%%%%%%%%%%%%%%%%%%%%%%%%%%%%%%%%%%%%%%%%%%%%%%%%%%%%%%%%%%%%%%%%%%%%%%%%%%%%%%%%%%%%%%%%%%%%%%%%%%%%%%    
    \begin{figure*}
        \centering
        \includegraphics[width=0.9\textwidth]{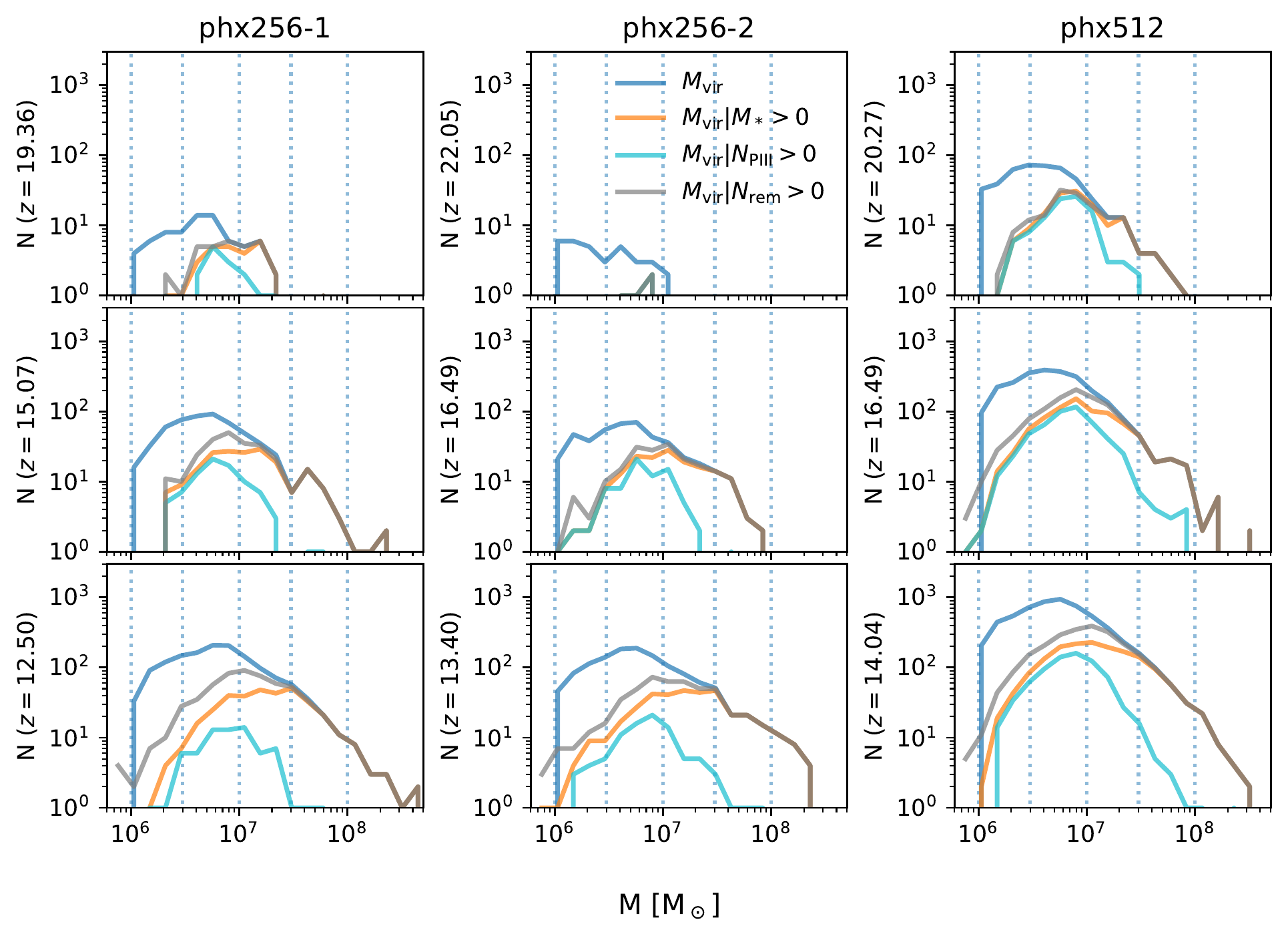}
        \caption{Halo distributions at different redshifts, as annotated on the vertical axis as raw counts of halos with the lowest panel taken from the final redshift of each simulation. PHX256-1,2 are shown at 80 Myr intervals, where PHX512 has 60 Myr intervals. $M_{vir}|M_*$ counts halos that have $M_* > 0$, $M_{vir}|N_{PIII}$ counts halos that contain active \piii stars, and $M_{vir}|N_{rem}$ counts those halos with \piii supernova remnants of any type.  At the final redshift, halos with mass $M_{vir} > 2\times 10^7$ M$_\odot$ all contain remnants, with no active \piii stars.  \piii stars found in halos with $M_{vir} < 2\times 10^6$ indicate that \piii formation may occur in under-resolved halos.}
        \label{fig:phx256-1_halo_occ}
    \end{figure*}
    Halo finding was performed using {\tt ROCKSTAR} \citep{behroozi2013}, requiring 50 DM particles per identified halo, however halo-based analyses are restricted to those halos with $\geq100$ particles.  Figure \ref{fig:phx256-1_halo_occ} shows the total halo mass function (HMF), the HMF of halos with non-zero stellar mass, the HMF of halos with active \piii stars, and the HMF of halos with \piii supernova remnants.     All simulations show that halos having $M_{vir} > 2\times 10^7~\msun$ are universally forming \pii stars at their final redshifts.  While some halos with $M_{vir} > 2\times 10^7~\msun$ are also forming \piii stars, the number is quickly diminishing beyond. Since we qualify a halo with $M_{vir} > 2.34\times 10^6~\msun$ as well-resolved, Figure \ref{fig:phx256-1_halo_occ} also shows that \piii star formation is occurring in under-resolved halos. 
%%%%%%%%%%%%%%%%%%%%%%%%%%%%%%%%%%%%%%%%%%%%%%%%%%%%%%%%%%%%%%%%%%%%%%%%%%%%%%%%%%%%%%%%%%%%%%%%%%%%%%%%%%%%%%%%%%%%%%%  
    \begin{figure}
        \centering
        \includegraphics[width=0.48\textwidth]{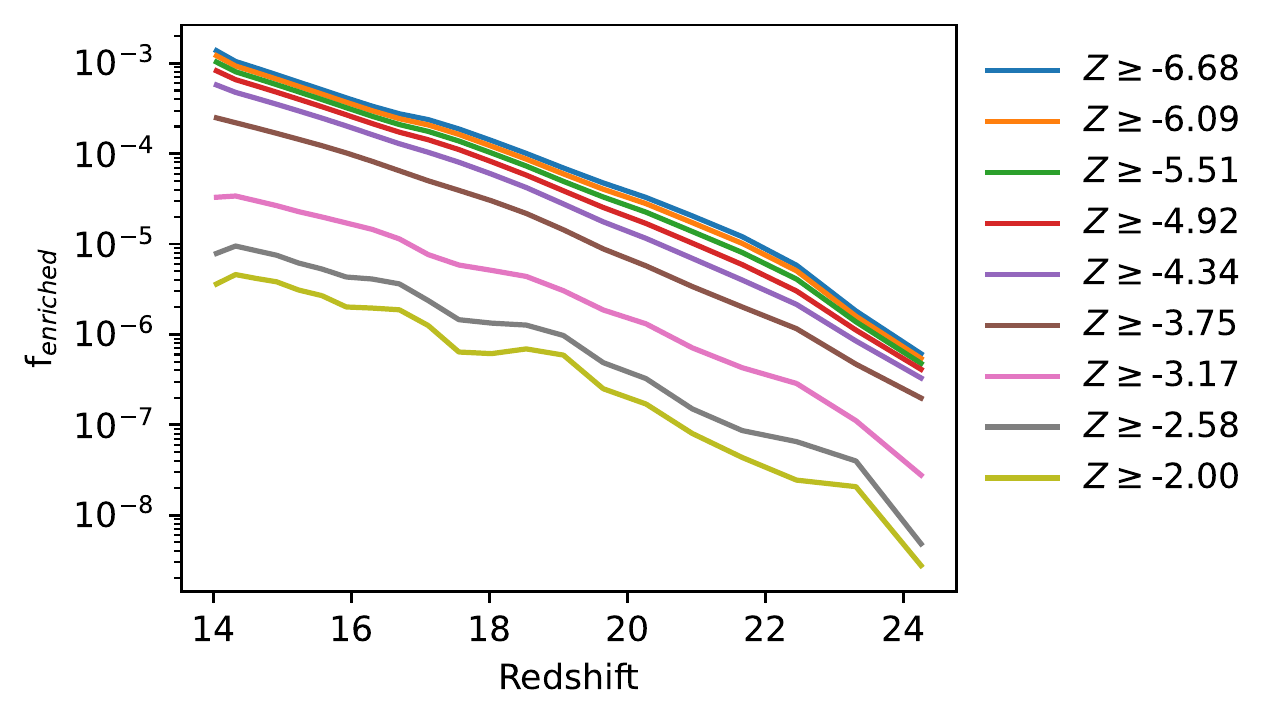}
        \caption{Fraction of gas enriched above varying $Z$ values in PHX512 looking back from $z=14.04$. Less than 0.1\% of the volume has been enriched to  $Z_{c}$ required for \pii star formation.}
        \label{fig:enriched_z-oden}
    \end{figure}
    
    Figure ~\ref{fig:enriched_z-oden} shows the fraction of gas that is enriched above varying cut-off $Z$ as a function of $z$ for $14.04 < z < 28$ in PHX512.  The IGM is largely unenriched at the lowest redshift, as indicated by the low fraction enriched to $Z\gtrsim -6.68$, which suggests that early \piii stellar feedback is not responsible for enrichment of the IGM .  In particular, potential \pii star forming gas with $Z \gtrsim -5.5$ accounts for $\sim0.08\%$ of the simulation volume.  We additionally analyzed the fraction of volume enriched by \piii versus \pii sources: if the \piii enriched fraction is $f_3$, and the fraction from all stars is $f_Z$, the quantity $\delta f = (f_Z - f_3)/f_Z$ shows the fraction of gas enriched by \pii sources.  We find that $\lesssim 2\%$ of enriched gas is enriched by \pii sources at the final redshift, emphasizing the importance of \piii chemical enrichment at high redshifts.
%%%%%%%%%%%%%%%%%%%%%%%%%%%%%%%%%%%%%%%%%%%%%%%%%%%%%%%%%%%%%%%%%%%%%%%%%%%%%%%%%%%%%%%%%%%%%%%%%%%%%%%%%%%%%%%%%%%%%%% 
    \begin{figure}
        \centering
        \includegraphics[width=0.48\textwidth]{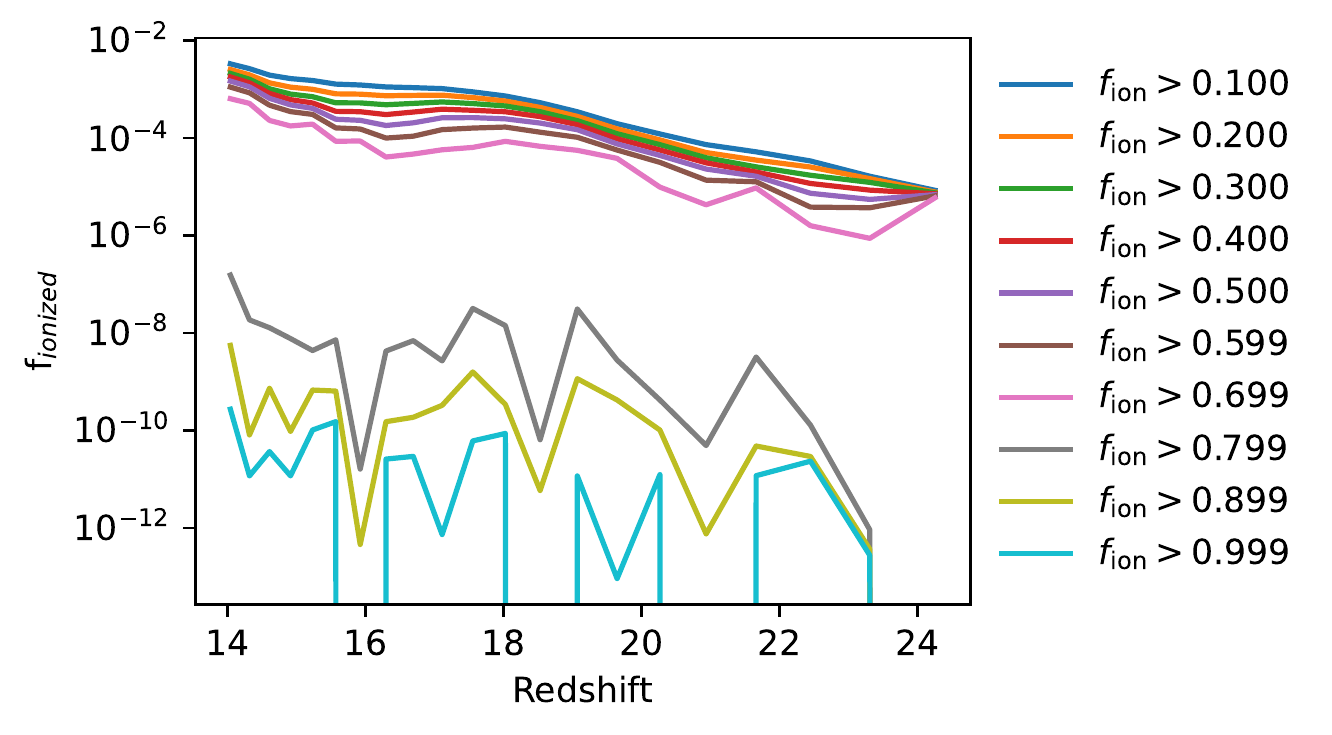}
        \caption{ Fraction of volume that is ionized to varying degrees, as measured by $f_{\rm ion} = n_{\rm H^+} / (n_{\rm H} + n_{\rm H^+})$.  At the final redshift, a negligible fraction of the volume is fully ionized ($f_{\rm ion} > 0.9$), indicating that IGM reionization has not begun to any substantial degree.}
        \label{fig:f_ionized}
    \end{figure}
    
    The ionization of the volume in PHX512 is shown in Figure \ref{fig:f_ionized}.  The ionized fraction is presented as $f_{\rm ion} = n_{\rm H^+}/ (n_{\rm H} + n_{\rm H^+})$; less than 0.5\% of the volume is ionized to $f_{\rm ion} > 0.3$ at the final redshift, indicating that the point-source radiation feedback from ionizing sources has not yet escaped the dense clumps of halo or galactic gas.  High $f_{\rm ion}\gtrsim 0.7$ likely results from ionizing radiation from \piii stars at these redshifts, with subsequent drops in $f_{\rm ion}$ resulting from recombination after the \piii main-sequence phase, as seen in the $f_{\rm ion} = \{0.799, 0.899, 0.999\}$ lines. Lower values in $f_{\rm ion}$ reflect the increasing volume affected by hydrodynamic heating and \piii supernova heated gas, as seen in the halo zoom-in temperature panel of Figure \ref{fig:sim_prj}. 
%%%%%%%%%%%%%%%%%%%%%%%%%%%%%%%%%%%%%%%%%%%%%%%%%%%%%%%%%%%%%%%%%%%%%%%%%%%%%%%%%%%%%%%%%%%%%%%%%%%%%%%%%%%%%%%%%%%%%%%    
    % COMBINE ALL SFR DATA TO SINGLE PLOT (ONE LINE FOR EACH SIM)
    \begin{figure}
        \centering
      
        \includegraphics[width=0.48\textwidth]{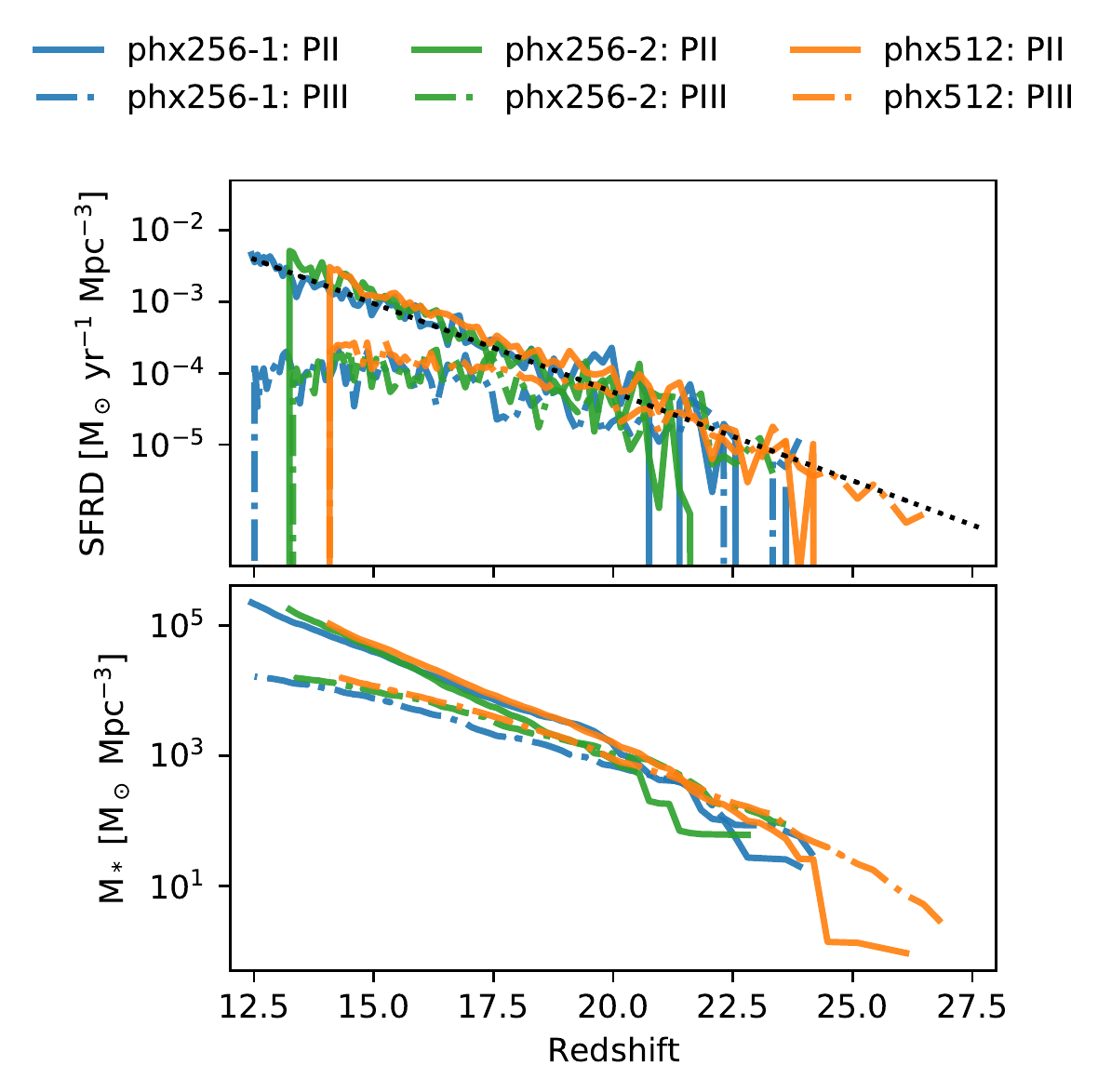}
        \label{fig:phx1_sfr_p2}
  
        \caption{ Star formation rate densities (top) and cumulative mass formed (bottom) for all Phoenix simulations. The era where \piii star formation happens at comparable rates to \pii is ended by $z\sim 23$, which also roughly corresponds to the era when more mass is formed in \pii stars than has existed in \piii.  The black dotted SFRD line shows an approximate fit to the \pii SFRD as $SFR_{\rm est}=5\exp(-z/1.75)$ $\msun$ yr$^{-1}$ Mpc$^{-3}$.}
        \label{fig:phxall_sfr}
    \end{figure}

    Star formation statistics are presented in Figure \ref{fig:phxall_sfr}.  The star formation rate density (SFRD) shows that there was comparable SFRD in \piii stars until $z\sim 19$, when the \pii SFRD surpasses it.  The \pii SFRD is fit by $SFR_{\rm est} = 5\exp(-z/1.75)$ $\msun$ yr$^{-1}$ Mpc$^{-3}$, shown on the figure. The mass panel shows the total formed mass of \piii stars, including the original mass of those that  have undergone a SN event.   The total mass in \pii stars surpasses the total mass formed in \piii stars by $z\sim 20$.  These panels also show that there is no clear delineation between \piii and \pii epoch in a volume-wide sense: \pii star formation commences $\lesssim 10$ Myr after the first \piii star is formed, and continues alongside \piii star formation for the entire duration of each simulation.

%%%%%%%%%%%%%%%%%%%%%%%%%%%%%%%%%%%%%%%%%%%%%%%%%%%%%%%%%%%%%%%%%%%%%%%%%%%%%%%%%%%%%%%%%%%%%%%%%%%%%%%%%%%%%%%%%%%%%%%
    \begin{figure}
        \centering
        \includegraphics[width=0.48\textwidth]{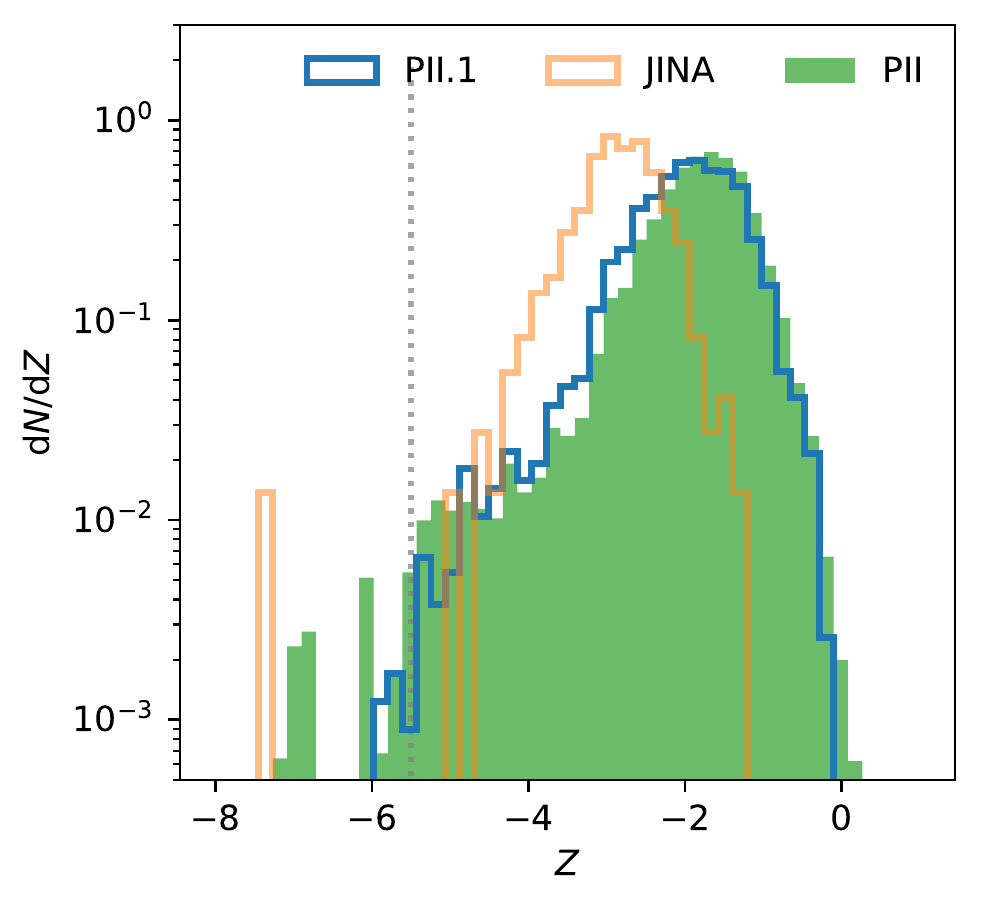}
        \caption{\pii metallicity distribution.  Solid: The MDF of all \pii stars at the final redshift across all simulations.  Blue, unfilled: The MDF of all \gii stars. The vertical line denotes the simulation $Z_c = -5.5$. Orange, unfilled: Sample MDF of observed low-metallicity stars from the JINA database \citep{abohalima2018}, including all stars from the Milky Way halo or dwarfs; the upper end ($Z > -2.5$) of this MDF is suppressed due to selection effects, and is not expected to resemble the simulation data. Notably, there is no obvious metallicity distinction between \gii clusters and those resulting from ongoing star formation.}
        \label{fig:pii_mdf}
    \end{figure}
    
    The metallicity distribution function (MDF) for \pii stars across all PHX simulations is presented in Figure \ref{fig:pii_mdf}.  The MDF of all \pii stars shows a large range of possible metallicities, with the most metal deficient cluster having $Z=-7.1$.  Although the simulation parameter for \pii cluster creation requires $Z>-5.5$ at the cell hosting cluster formation, the mass-averaged metallicity of the gas that contributed to star formation allows these low-$Z$ clusters to exist.  The highest $Z$ cluster has $Z=0.57$, with mean {$\langle Z\rangle = -1.89 $}.  The unfilled blue curve shows only \gii stars, with mean {$\langle Z\rangle = -2.00$}. The similarity of the \gii MDF to the total \pii MDF is striking, suggesting that stellar metallicity is not a good proxy for its age.

%%%%%%%%%%%%%%%%%%%%%%%%%%%%%%%%%%%%%%%%%%%%%%%%%%%%%%%%%%%%%%%%%%%%%%%%%%%%%%%%%%%%%%%%%%%%%%%%%%%%%%%%%%%%%%%%%%%%%%%
%%%%%%%%%%%%%%%%%%%%%%%%%%%%%%%%%%%%%%%%%%%%%%%%%%%%%%%%%%%%%%%%%%%%%%%%%%%%%%%%%%%%%%%%%%%%%%%%%%%%%%%%%%%%%%%%%%%%%%%
%%%%%%%%%%%%%%%%%%%%%%%%%%%%%%%%%%%%%%%%%%%%%%%%%%%%%%%%%%%%%%%%%%%%%%%%%%%%%%%%%%%%%%%%%%%%%%%%%%%%%%%%%%%%%%%%%%%%%%%

\section{Analysis: The first stars and the second generation}
    \label{sec:analysis}
    To study the origin of \gii star formation, we take two primary frames of reference: A) we analyze \piii star forming regions, studying the evolution of the region as it leads to \pii star formation, and B) we examine the region about the \gii cluster immediately after formation, to study the events that immediately contributed to its formation.  Each frame of reference uses separate analyses of the simulations.  
        \subsection{Method A}
        \label{sec:A}
            \piii star formation within the PHX suite is clustered, which is not unexpected (e.g., \citealt{stacy2010}): if we follow a single \piii star forming region as defined in this section, we find that there are up to 167 individual \piii star formations per region.  Although clustered, there are not enough individual stars to qualify as a star cluster in the canonical sense: we will therefore refer to them as \systems to avoid ambiguity with modern or \pii star clusters, but in analogy to modern O-B associations.
            
            To examine the simulations from the perspective of \piii stars and define the extent of a \system, we iterate each output of the simulation to find new \piii star particles that formed within 200 kyr by iterating dark matter halos, searching for new particles within $3 R_{vir}$ of the center of the halo.  When found, and if that star is only accompanied by coeval \piii star formation (i.e., other new \piii stars only formed within 200 kyr, no supernova remnants, black holes, or \pii stars within the radius), we form a sphere to measure mean metallicity from new \piii stars ($\bar Z_{III}$) and H$^+$ fraction($\bar f_{H^+}$), using the analysis software {\tt yt}~\citep{turk2011}. Starting at $R = 250$ pc, the average value for a field, $\bar X$, is measured. If $\bar X$ is not less than some critical value, $R$ is increased to $R_{i+1} = 1.1 R_i$, and a new average is taken.  This procedure is repeated until the value of $\bar X$ is below our chosen critical values: ${-5}$ for $\bar Z$ and $0.05$ for $\bar f_{H+}$.  The ``edge" of the region is then defined by this final radius.  The final product of this analysis is an effective radius as function of time for the $Z_{PIII}$ and $f_{H+}$ variables
        \subsection{Method B}
        \label{sec:B}
            From the perspective of \pii clusters, we again iterate through simulation outputs to identify \pii star formation events.  When a new \pii star particle is formed, and it occurs in gas enriched by only \piii stars (i.e., the metallicity from \pii stars, $Z_2$, meets the criteria $\bar Z_2 < -6$ within 500 pc), we evaluate a sphere centered on that star with $r= 200$ comoving kpc.  Within this sphere, we connect any \piii supernova remnant to the \pii cluster via a ray. We then verify that the ray has $Z > Z_c$ for all cells it intersects and requiring that distance between particles, $d$, satisfies $d \leq vt$ for $v = 100 km/s$ SN remnant expansion speed, and $t$, the time between SN and \pii formation. $d\leq vt$ implies that metals from the \piii star could feasibly have reached the forming star cluster and acts as a filter for overlapping metal clouds from separate \systems. If these criteria are satisfied, then that \piii event is ``connected" to the \pii formation and is considered to be part of the same ``metal system". The extremely fine time resolution between outputs ensures that \piii events connected to the formed \pii cluster were connected at the time of formation ($\leq 200$ kyr prior).
%%%%%%%%%%%%%%%%%%%%%%%%%%%%%%%%%%%%%%%%%%%%%%%%%%%%%%%%%%%%%%%%%%%%%%%%%%%%%%%%%%%%%%%%%%%%%%%%%%%%%%%%%%%%%%%%%%%%%%%
    \subsection{The \piii frame}
    \label{sec:p3regions}
    % add vertical line for mean of each type in each panel

    \begin{figure}
    \centering

    \subfloat[$Z_{III}$ Radius]{
        \includegraphics[width=0.44\textwidth]{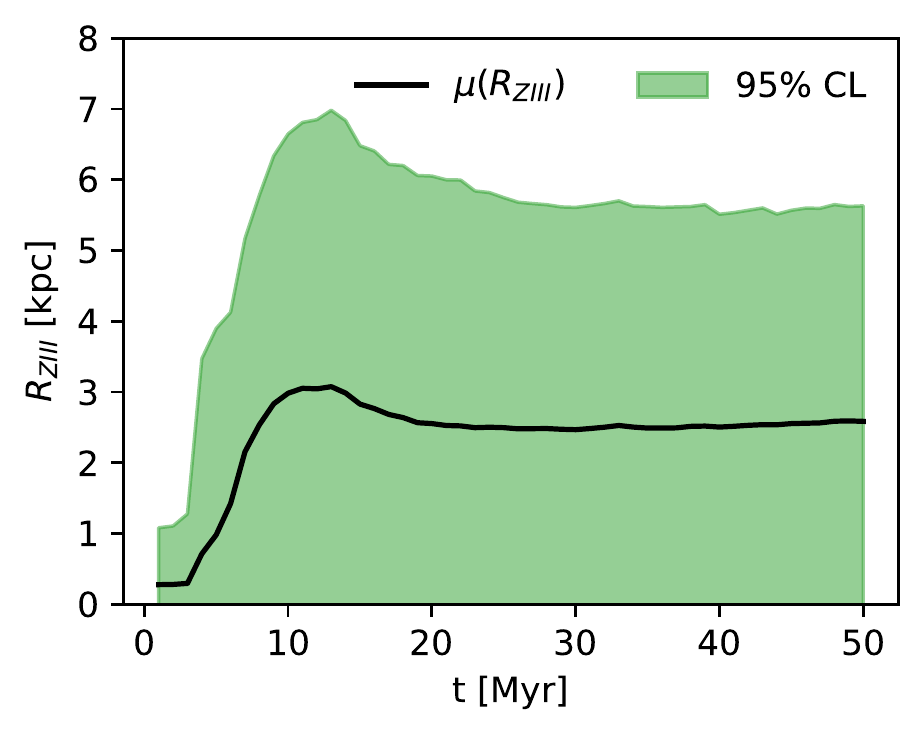}
        \label{fig:p3region-Z}}\\
    \subfloat[H$^+$ Radius]{
                \includegraphics[width=0.44\textwidth]{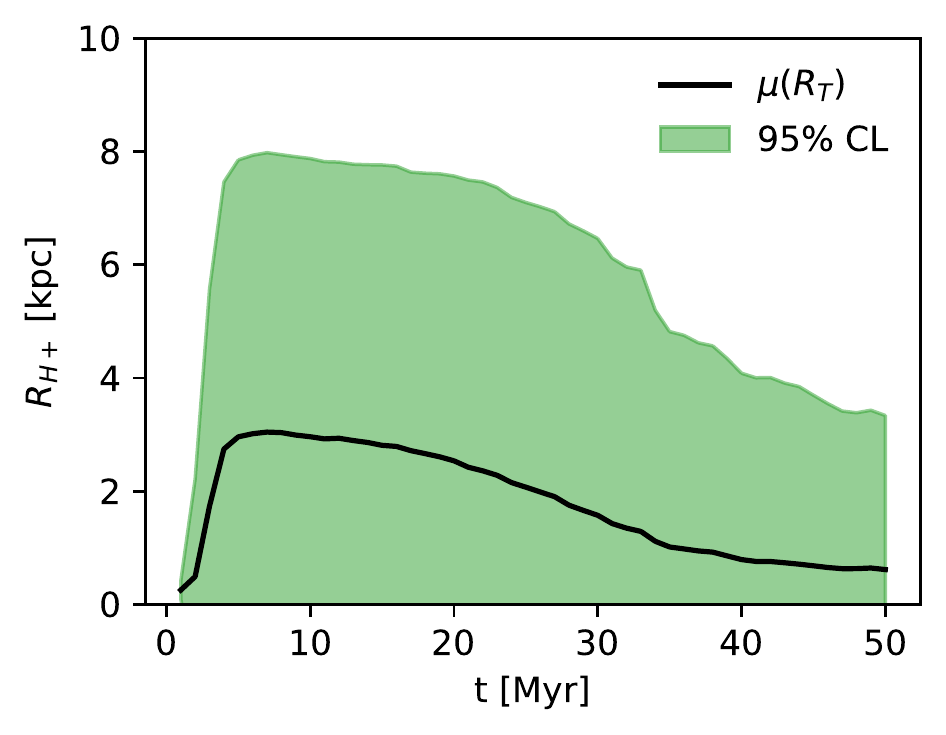}
                \label{fig:p3region-temp}
                    }
    \caption{\piii star forming region characteristics spanning the first 50 Myr past the first particles' formation.  \ref{fig:p3region-Z} shows the radii of the \piii metal cloud, as determined in Section \ref{sec:A}.  The volume-weighted average ($\mu$) is plotted in black in black, and 95\% confidence level in green.  \ref{fig:p3region-temp} shows the same for the H$^+$ field.} 
    \label{fig:p3regions}
    \end{figure}

    The time evolution of metal and ionized radii of \systems is presented in Figures~\ref{fig:p3region-Z} and \ref{fig:p3region-temp} respectively; we also indicate the 95\% confidence limit on the distribution of measured values, i.e., the bands represent 95\% of the distribution of measured radii. After the initial formation, both radii take on the minimum 0.25 kpc, reflecting that $Z_{III}$ is sourced from supernovae that have not occurred yet, and that H ionization requires time to first ionize the dense cloud that the particle formed within.  The average radius for $Z_{III}$ ($R_{III}$) is maximum at 9-12 Myr after the first formation: This is the time-frame expected for most \piii supernovae of all types to have reached their main sequence endpoint if we are observing a coeval system of stars.  The reduction in $R_{III}$ likely reflects gravitational collapse, suppported at later times by increased temperatures and the feedback from \pii stars.  The radius of H ionization, $R_{H+}$, increases quickly to a maximum average within 5 Myr.  The reduction beyond this point reflects recombination after the most massive (and ionizing) \piii sources of radiation have extinguished.
%%%%%%%%%%%%%%%%%%%%%%%%%%%%%%%%%%%%%%%%%%%%%%%%%%%%%%%%%%%%%%%%%%%%%%%%%%%%%%%%%%%%%%%%%%%%%%%%%%%%%%%%%%%%%%%%%%%%%%%
    \begin{figure}
    \centering

    \subfloat[\piii SFR]{
        \includegraphics[width=0.49\textwidth]{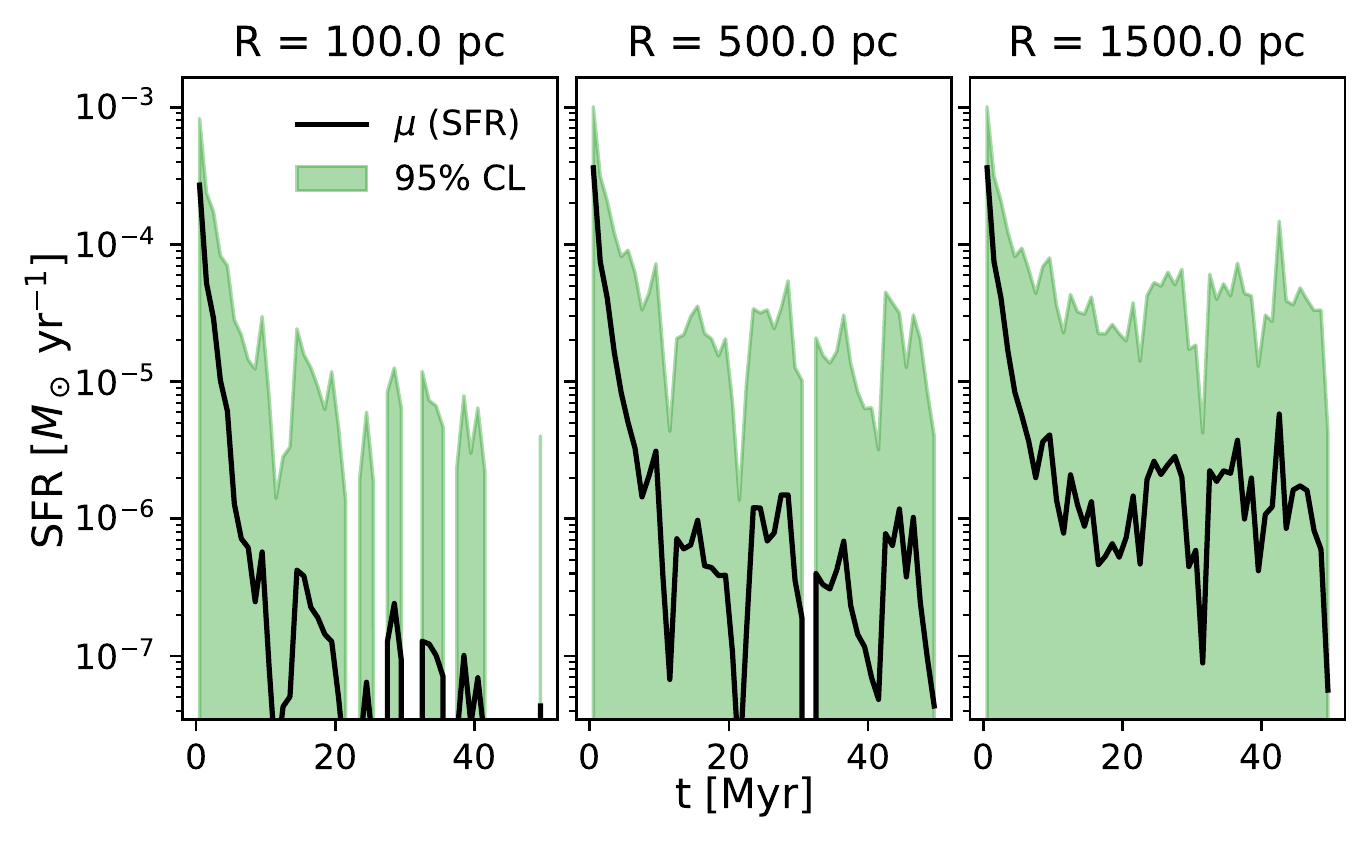}
        \label{fig:p3_sfr}}\\
    \subfloat[\pii SFR]{
                \includegraphics[width=0.49\textwidth]{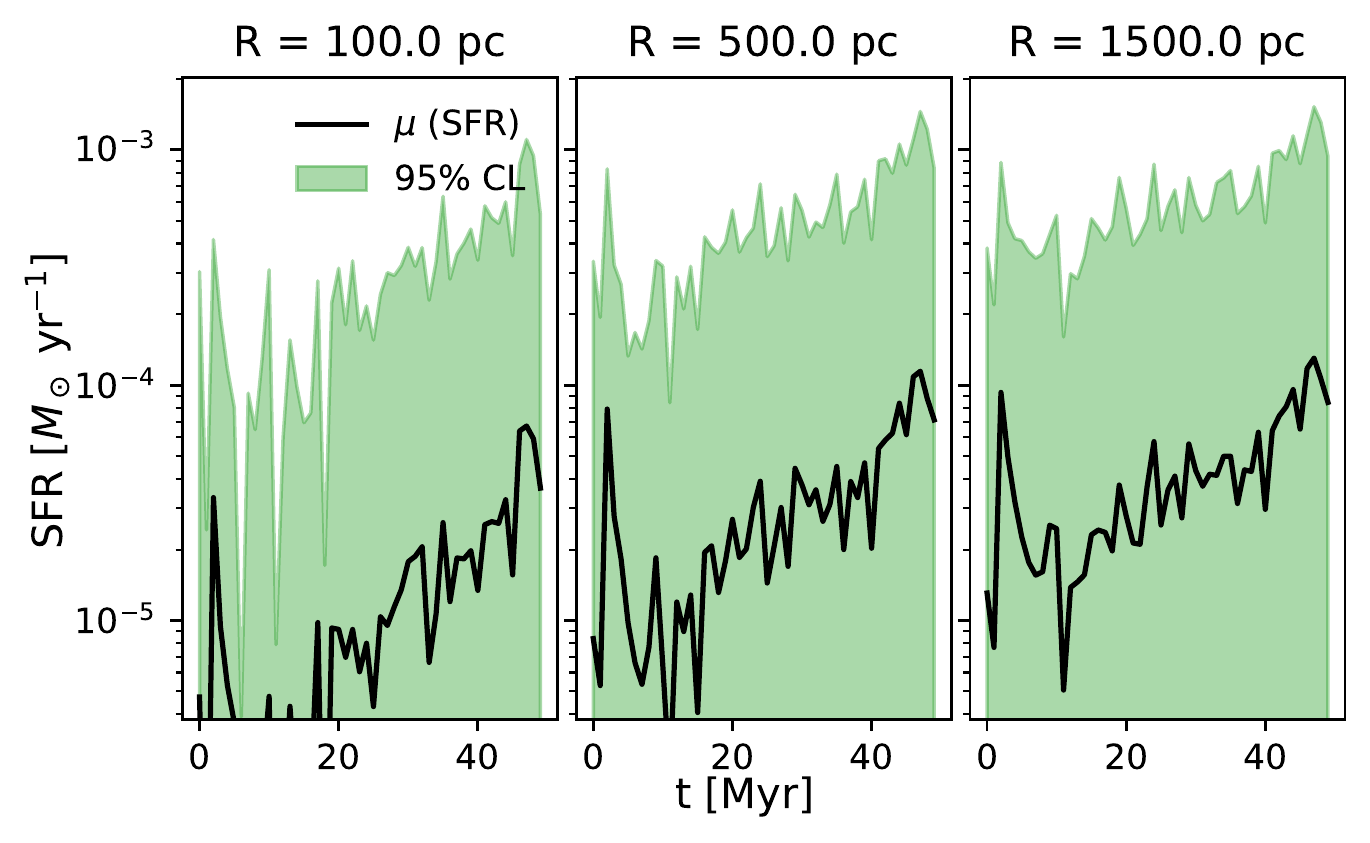}
                \label{fig:p2_sfr}
                    }
    \caption{Star formation rates considering varying radii from the first \piii particle to form within the region.  Changes in the SFR with increasing radius indicate star formation nearby the original particle.} 
    \label{fig:region_SFRs}
    \end{figure}
    
    In Figure~\ref{fig:region_SFRs}, we present the averaged SFR of each region with 95\% confidence level for \npiii ~distinct primordial star forming regions.  We present several SFRs within spheres of radius $R = \{100,500,1500\}$ pc from the first particle to form.  The \piii SFR quickly drops to negligible levels within 10 Myr of the first star, although nearby systems are indicated by the higher SFR at larger radii.  As expected, the \pii SFR is inversely related to the \piii rate, starting negligible and increasing to $\sim 8\times 10^{-5} \msun$/yr by 50 Myr.  There is also \pii star formation in close proximity to new \systems, evident by observing the non-zero \pii SFR at $R\leq 1500$ pc even at $t\sim 0$.  The early spike in \pii SFR at $t\sim 2$ is due to triggered star formation in the aftermath of the earliest \piii supernovae.
%%%%%%%%%%%%%%%%%%%%%%%%%%%%%%%%%%%%%%%%%%%%%%%%%%%%%%%%%%%%%%%%%%%%%%%%%%%%%%%%%%%%%%%%%%%%%%%%%%%%%%%%%%%%%%%%%%%%%%%

%%%%%%%%%%%%%%%%%%%%%%%%%%%%%%%%%%%%%%%%%%%%%%%%%%%%%%%%%%%%%%%%%%%%%%%%%%%%%%%%%%%%%%%%%%%%%%%%%%%%%%%%%%%%%%%%%%%%%%%
\subsection{The \gii frame}

    \begin{table*}
        \centering
        \caption{Progenitors of \gii cluster formation}
        \begin{tabular}{|c|c|c|c|c|c|c|}\hline\hline
            Configuration & $N_{\rm II}$ & $f_{\rm II}$ & $\langle Z\rangle$ & $\langle R_{f}\rangle$ [pc] & $N_{\rm III}$ & $M_{\rm III}[\msun]$\\\hline
            SN & 31 & 0.007 & $-1.73\pm 0.49$ & $92.41\pm 107.37 $ & $ 3.2\pm 2.6$ & $ 48.7\pm 38.5$\\
            HN & 320 & 0.078 & $-2.04\pm 0.68$ & $77.26\pm 88.65 $ & $ 3.1\pm 2.2$ & $ 92.2\pm 60.3$\\
            PISN & 616 & 0.120 & $-2.15\pm 0.67$ & $30.51\pm 30.28 $ & $ 1.1\pm 0.4$ & $ 218.8\pm 83.3$\\
            SN-HN & 896 & 0.201 & $-2.02\pm 0.63$ & $201.83\pm 339.92 $ & $ 13.7\pm 9.9$ & $ 331.2\pm 257.6$\\
            SN-PISN & 17 & 0.002 & $-1.79\pm 0.53$ & $132.34\pm 167.77 $ & $ 3.7\pm 1.6$ & $ 236.7\pm 57.2$\\
            HN-PISN & 220 & 0.044 & $-1.97\pm 0.58$ & $114.88\pm 243.31 $ & $ 4.1\pm 2.7$ & $ 337.9\pm 158.4$\\
            SN-HN-PISN & 2446 & 0.547 & $-2.01\pm 0.61$ & $407.48\pm 591.27 $ & $ 20.2\pm 13.4$ & $ 775.0\pm 443.2$\\\hline\hline
        \end{tabular}
        \tablecomments{Characteristics of \ngii \gii star clusters given the connected \piii events within the region.  We categorize each region containing a new \pii star by the type of \piii supernovae it contained: SN, HN, PISN, or combinations thereof.  For each configuration, we present the number of \gii stars formed ($N_{\rm II}$, the fraction of \gii stellar mass generated ($f_{\rm II}$), the mean metallicity ($\langle Z\rangle$), the mean radius from progenitor to forming cluster ($R_f$), the mean number of progenitors in the region ($N_{\rm III}$), and the mean total progenitor mass ($M_{\rm III}$). Note that  78.7\% (by mass) of \gii stars have progenitors of multiple types.
        }
        \label{tab:2g_stats}
    \end{table*}
    Generated using method B, we present the statistics of \piii stars connected to {\ngii} \gii star clusters in Table \ref{tab:2g_stats}. The region surrounding the \gii star particle is categorized based on the type of \piii progenitors it contains: SNe, HNe, PISNe, or any combination of the three.  There are several immediate observations worth noting: there is no obvious correlation between the mean metallicity of \gii star clusters given the progenitor configuration; the mass of the \piii stars that contributed to formation is highly variable; and finally, despite having $M_c=20 \msun$ in the IMF, $12\%$ of \gii stars were enriched by an average of $N_{\rm III} = 1.1$ PISN, while other \gii stars are enriched by $N_{\rm III}> 3$ \piii progenitors.  The average distance from progenitor to \gii cluster is maximized if all types of \piii progenitors are present and connected to the \gii star. Outside the single-type cases of the SNe and HNe, the average \gii star is connected to $> 200~\msun$ of \piii progenitors, with some \gii stars connecting to $> 1000~\msun$ of \piii supernova generating stars.  It is worth noting that the lack of correlation between cluster metallicity and progenitor configuration may be an artifact of the star formation algorithm: since the \pii clusters take on the mass-averaged metallicity of the cold gas that formed them, we lose the ability to track extreme cases that may occur with higher resolution and modelling \pii stars as single stars.  That said, within the framework of the PHX simulations, the highest metallicity star clusters $(Z = -1.73)$ result from regions with only SNe, while the lowest metallicity $(Z=-2.15)$ have single PISNe progenitors.

\section{An Interpretable Regression Model of primordial stars' Influence}
\label{sec:model}
    In this section, we develop and describe an interpretable model designed to predict  the range of influence of \systems.  Using data generated via method B in Section~\ref{sec:B}, we will create a model to learn the extent of primordial metals from a \system based on it's composition. To translate the continuum of available \piii masses and creation times, we transform the stellar information into a simple series of features as binned mass and creation times.  The mass features (${\bf X}_M$) are defined by edges $\{1,11,20,40,100,140,200,260,300\}~\msun$, and creation time bins (${\bf X}_C$) have edges \{$\delta t$, 2$\delta t$, ...,$t_{\rm final}$\} Myr, with the time between bin edges ($\delta t$) and the final time ($t_{\rm final}$) being tunable hyperparameters.  We split samples in a spatial sense to prevent information leaking between training and testing splits:  if the center of the region (${\bf r}$) at the first star's formation is ${\bf r} > \{0.4, 0.4, 0.5\}\times L_{box}$ relative to the simulation volume, then the sample is assigned to the test split, ${\bf r} < \{0.6, 0.6,0.5\}\times L_{box}$ is assigned to validation, while all others are assigned to training.  This splitting of data yields 2,273 training, 722 validation, and 318 testing samples.

    \subsection{Data}
    \label{sec:data}
    \begin{figure}
        \centering
        \includegraphics[width=0.47\textwidth]{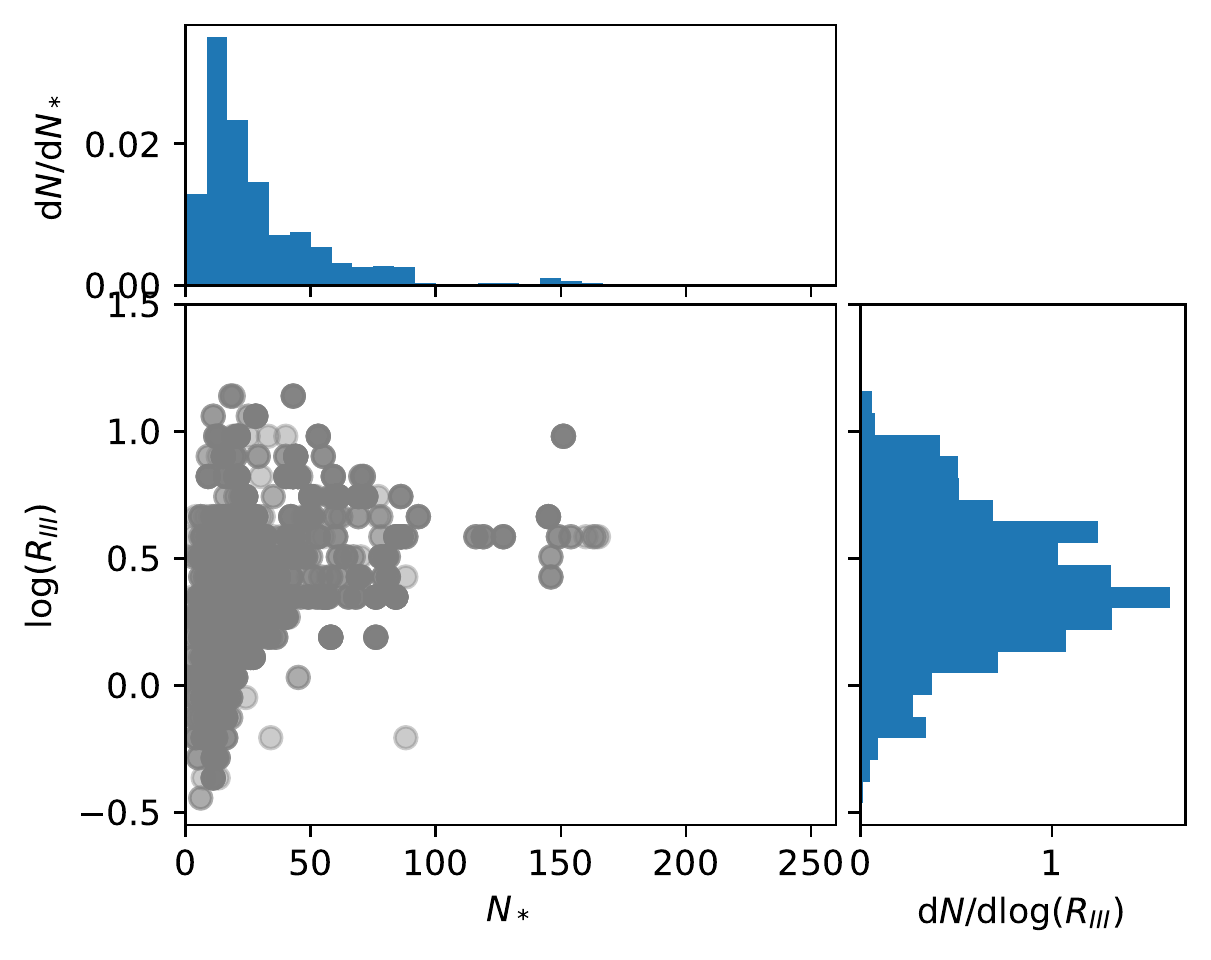}
        \caption{Training data for 18 Myr after first \piii formation.  Histograms show PDFs of $N_*$ (top) and $R_{III}$ (side).}
        \label{fig:traindata}
    \end{figure}
     Figure \ref{fig:traindata} shows $\log (R_{III})$ according to $N_*$ within the region for the training examples at $t_{\rm final} = 18$ Myr.  The side histogram shows a PDF of $\log (R_{III})$; while not perfectly log-normal, the log($R_{III})$ distribution is distinctly more evenly distributed about the peak ($\log (R_{III})\simeq 0.4$) than without the logarithm:  we therefore make our predictions on $\log (R_{III})$, as regression is more successful at modelling scatter from a mean than long tails of distributions. We can also observe a huge variation in the number of \piii stars contained within $R_{III}$.  While the sphere defined by $R_{III}$ undoubtedly contains multiple star systems in some cases, we can observe $1 < N_* < 167$ within that radii.  For each $t_{\rm final}$, we define $\bar N_* = \mu(\log N_*)$, and $\sigma_N = \sigma(\log N_*)$\footnote{$\mu$ and $\sigma$ represent the mean and standard deviation of the distribution, respectively}, then for 18 Myr, $\bar N_* = 1.337$ and $\sigma_N = 0.374$.  In addition, while we can observe a general trend that increasing $N_*$ increases $R_{III}$, however this relationship is by no means linear in $N_*$, and is not single-valued, i.e., $N_* = 10$ could lead to many values in the range $-0.5 < \log (R_{III}) < 0.8$.

    \subsection{Model and hyperparameters}
    To generate an interpretable model, we use a simple linear regression model, where we wish to minimize the error
        \begin{equation}
        L = \frac{1}{2}({\bf Y}-\hat {\bf Y})^2,
        \label{eqn:loss}
    \end{equation}
    with ground-truth radii $\bf Y$ and predicted radii $\hat {\bf Y}$ represented as vectors. The prediction is simply solved by $\hat {\bf Y} = {\bf X} \cdot {\bf w}$ where ${\bf X} \equiv \{1\}\oplus {\bf X}_M \oplus {\bf X}_C$\footnote{The concatenation operation is represented by $\oplus$} and ${\bf w}$ are the weights to learn.  The solution we seek is therefore $dL/d{\bf w} = 0$, which is solved by 
    \begin{equation}
        {\bf w} = (\bf X\bf X^T)^{-1}\bf X^T\bf Y.
    \end{equation}
    This regression only admits linear behavior, which as outlined in Section \ref{sec:data}, is not a reasonable expectation in this case.  The non-linearity is not only influenced by $N_*$, but also their varying lifetimes, explosion energies, metal yields, birth times, and supernovae times.  Instead of abandoning the linear model for a less interpretable deep learning architecture, we split the training samples into subsets based on $N_*$ and apply the linear model to each subset: we use the samples $\textbf{X}_n$ to train only the model $M_n$ according to 
    \begin{equation}
        M_n = \begin{cases}
                M_0, & \textbf{X} | \log N_* < \bar{N}_*-\sigma_N \\
                M_1, & \textbf{X} | \bar{N}_*-\sigma_N < \log N_* < \bar{N}_* \\
                M_2, & \textbf{X} | \bar{N}_* < \log{N}_* < \bar{N}_*+\sigma_N \\
                M_3, & \textbf{X} | \log N_* > \bar{N}_*+\sigma_N.
        \end{cases}
    \end{equation}
    In this way, we have piece-wise defined models specifically for small systems with few stars, average systems, and highly populated systems.
    
    To evaluate the model predicted radii, we employ two metrics: we use the $R_2$ score for $N$ samples, given by 
    \begin{equation}
        R_2({\bf Y}, \hat {\bf Y}) = 1 - \frac{1}{N} \frac{\sum_{i=1}^N y_i - \hat y_i}{\sum_{i=1}^N y_i - \bar y},
    \end{equation}
    for $\bar y = 1/N \sum_{i=1}^N y_i$ and $y_i$ are the individual components of {\bf Y}.  The $R_2$ score is informative in comparing the quality of the model as compared to simply predicting the mean value for of {\bf Y}; 1 is a perfect $R_2$ score, $R_2=0$ indicates predicting $\hat y =\bar y$ for all inputs, and $R_2 < 0 $ indicates arbitrarily worse performance.  In addition to the individual predictions, we would also like the PDF of predicted radii to match that of the ground truth radii. We define the PDF of {\bf Y} as $P_y$, $\hat {\bf Y}$ as $P_{\hat y}$, and the Kullback-Leibler divergence $D(P|Q) = P\log (P|Q)$ for probability distributions $P$ and $Q$, to compare the PDFs using the Jensen-Shannon distance given by
    \begin{equation}
        J(P_y, P_{\hat y}) = \sqrt{\frac{D(P_y|\bar P_y) + D(P_{\hat y}|\bar P_y)}{2}},
    \end{equation}
    where $\bar P_x$ represents the average of the distribution $P_x$.  $J = 0$ represents two identical distributions, with higher values indicating mismatches in the PDFs.  Although not used in the initial evaluation of models, we include reports on the average $L_2$ distance, $\bar L_2 = 1/N \sum_{i=0}^N (y_i - \hat y_i)^2$, and $L_1$ distance as $\bar L_1 = 1/N \sum_{i=0}^N |y_i - \hat y_i|$.

    \subsection{Model results}
        \begin{table}[]
        \centering
        \caption{Testing dataset performance varying modelled time.}
        \begin{tabular}{|c|c|c|c|c|}
                \hline\hline
                Time &  $R_2$ & J     & $\bar L_2$   & $\bar L_1$ \\\hline
                7 Myr & 0.584 & 0.188 & 0.051 & 0.178\\
                8 Myr & 0.584 & 0.189 & 0.036 & 0.142\\
                9 Myr & 0.529 & 0.180 & 0.034 & 0.141\\
                10 Myr & 0.351 & 0.192 & 0.048 & 0.169\\
                11 Myr & 0.488 & 0.232 & 0.041 & 0.158\\
                12 Myr & 0.458 & 0.194 & 0.041 & 0.159\\
                13 Myr & 0.400 & 0.184 & 0.037 & 0.155\\
                14 Myr & 0.407 & 0.196 & 0.041 & 0.169\\
                15 Myr & 0.382 & 0.223 & 0.038 & 0.156\\
                16 Myr & 0.480 & 0.136 & 0.034 & 0.151\\
                17 Myr & 0.439 & 0.150 & 0.031 & 0.141\\
                18 Myr & 0.347 & 0.162 & 0.035 & 0.154\\
                19 Myr & 0.377 & 0.208 & 0.040 & 0.155\\

                \hline\hline
        \end{tabular}
        \tablecomments{These models use $\delta t = 6$ Myr.  Attempts to model times $<7$ Myr tends to produce lower $R_2$ and higher $J$, likely due to the strong dependence on dynamic evolution and hydrodynamic state at early times.}        
        \label{tab:modelresults}
    \end{table}
    \begin{table*}[]
        \centering
        \caption{Final parameters for linear regression models with 6 Myr time bin width}
        \begin{tabular}{|c|c|}
        \hline
            $t$ & ${\bf w}(M_0,M_1,M_2, M_3)$\\\hline\hline
            % 6 Myr &             &   &       &           &       \\
           \multirow{4}{*}{8 Myr}   &\{-0.43925, -0.04462, -0.02925, 0.07348, 0.02791, -0.04558, 0.26391, 0.42991, 0.13344, 0.03020\} \\
                                    &\{-0.05569, -0.02747, -0.01467, 0.03098, -0.01611, 0.00637, 0.19142, 0.32945, 0.07387, 0.00929\} \\
                                    &\{0.32175, -0.01745, -0.00265, 0.01291, -0.00351, -0.02874, 0.03616, 0.17712, 0.05762, 0.00086\}  \\
                                    &\{0.45087, 0.00059, -0.00346, 0.00481, -0.00010, -0.00794, -0.01343, 0.06214, -0.01298, 0.00175\} \\\hline

            \multirow{4}{*}{18 Myr}   & \{-0.19197, -0.10310, -0.06437, -0.02276, -0.02599, -0.10988, 0.15441, 0.28160, -0.07763, 0.07852, 0.12275\}  \\
                                        &\{-0.01883, -0.02943, 0.00425, 0.01444, -0.02386, 0.01085, 0.14379, 0.24969, 0.03048, 0.01509, 0.01826\}  \\
                                        &\{0.22457, -0.01837, 0.00044, 0.01954, -0.00042, -0.04703, 0.05068, 0.16201, -0.05797, 0.00048, -0.00245\} \\
                                        &\{0.52005, -0.02404, 0.01088, 0.00142, -0.01067, -0.00405, 0.01206, 0.02492, 0.02561, 0.00045, 0.00014\} \\
             \hline\hline

        \end{tabular}
        \tablecomments{ These exemplary weights are only for 8 Myr and 18 Myr, however the data table with full machine precision for all models presented in Table \ref{tab:modelresults} will be available at www.rensimlab.github.io. These models use $\bar N_* = 1.311, 1.335$, and $\sigma_N = 0.352, 0.349$ for 8 Myr and 18 Myr models respectively.}        
        \label{tab:trainedweights}
    \end{table*}
        \begin{figure}
            \centering
            \includegraphics[width=0.45\textwidth]{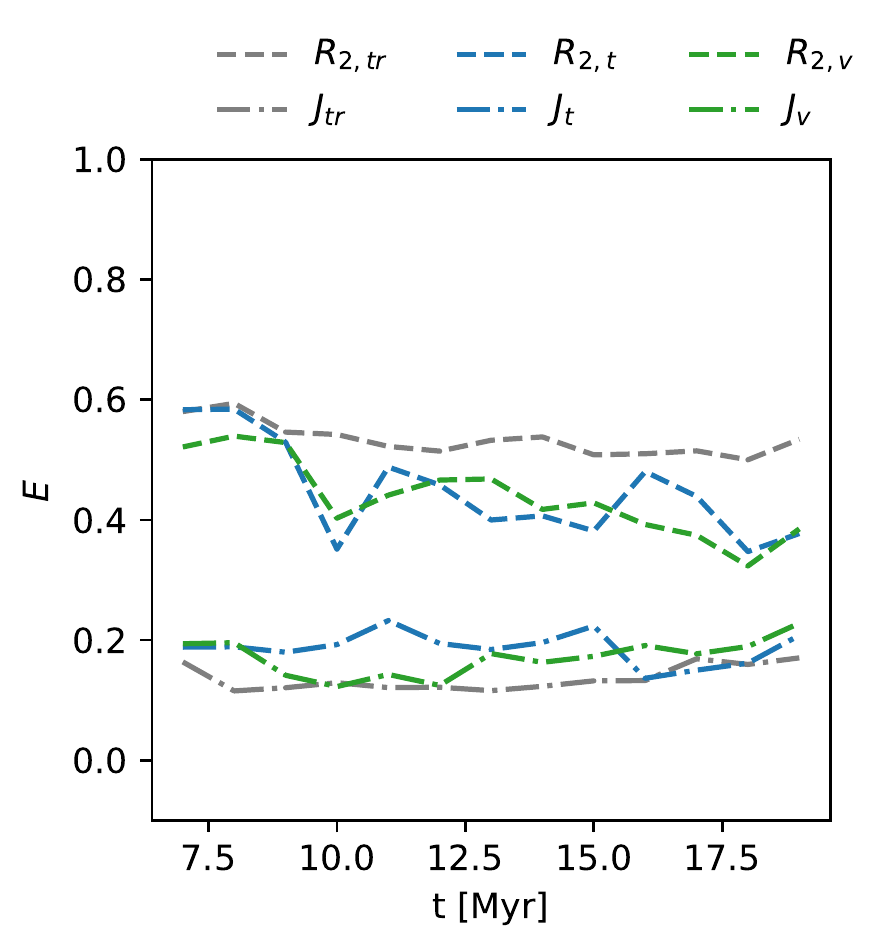}
            \caption{Error ($E$) as measured by $J$ (dash-dot lines) and $R_2$ (dashed lines) on the training (grey), validation (blue) and testing (green) data splits. This method is generally capable if the modelled time is $\geq 7$ Myr, but struggles by these metrics if $t < 5$ Myr, as the early dynamic evolution and dependence on specific hydrodynamic state of the system is making the modelling task more difficult at those early times.}
            \label{fig:jandr2}
        \end{figure}
    
        The only true hyperparameter of this model is $\delta t$, the time width of the ${\bf X}_C$ bins. We tested several widths, and found no significant improvement beyond the inclusion of two time features: early coeval formation, and all other formation, so all results here use $\delta t = 6$ Myr.  $J$ and $R_2$ as functions of predicted time in the all datasets are shown in Figure \ref{fig:jandr2}, while tabulated quantification of errors are presented in Table \ref{tab:modelresults}.  All times have $0.12 < J < 0.25$, while $R_2$ has more variation with $0.3 < R_2 < 0.6$.  These results show that while the model may struggle, e.g., with the lowest $R_2$ score at $T=18$ Myr, to reproduce exact predictions, it does well at reproducing the distribution of possible radii.  Based on Table \ref{tab:modelresults}, if we take the ``best" performing model as the one that minimizes $J$, $\bar L_1$, and $\bar L_2$ while maximizing $R_2$, we find that models with $t_{\rm final} = \{8, 16\}$ Myr are the best performing, while $t_{\rm final} = \{10, 15, 18\}$ have the worst performance.
        
        The final weights of the exemplary $t_{\rm final} = \{8,18\}$ Myr models are presented in Table \ref{tab:trainedweights}.  These weights represent the entirety of the trained model.  Due to their simplicity, erroneous predictions could be traced through the model to find the offending weight and determine why the model made such an error very easily.

%%%%%%%%%%%%%%%%%%%%%%%%%%%%%%%%%%%%%%%%%%%%%%%%%%%%%%%%%%%%%%%%%%%%%%%%%%%%%%%%%%%%%%%%%%%%%%%%%%%%%%%%%%%%%%%%%%%%%%%
        \begin{figure*}
        \centering
        \subfloat[8 Myr]{
        \includegraphics[width=0.49\textwidth]{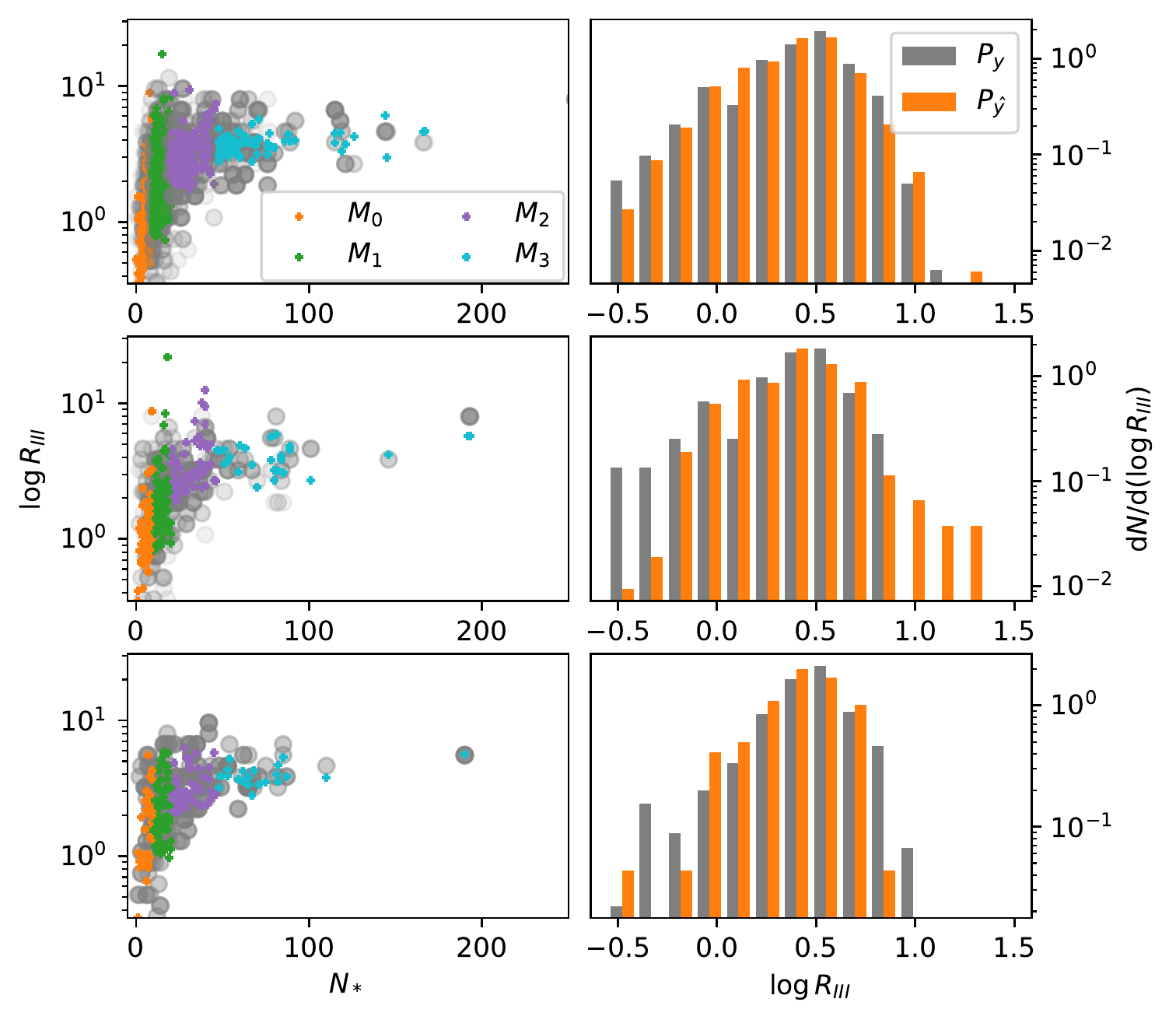}
        \label{fig:8x4_regmodel}}
        \subfloat[18 Myr]{
        \includegraphics[width=0.49\textwidth]{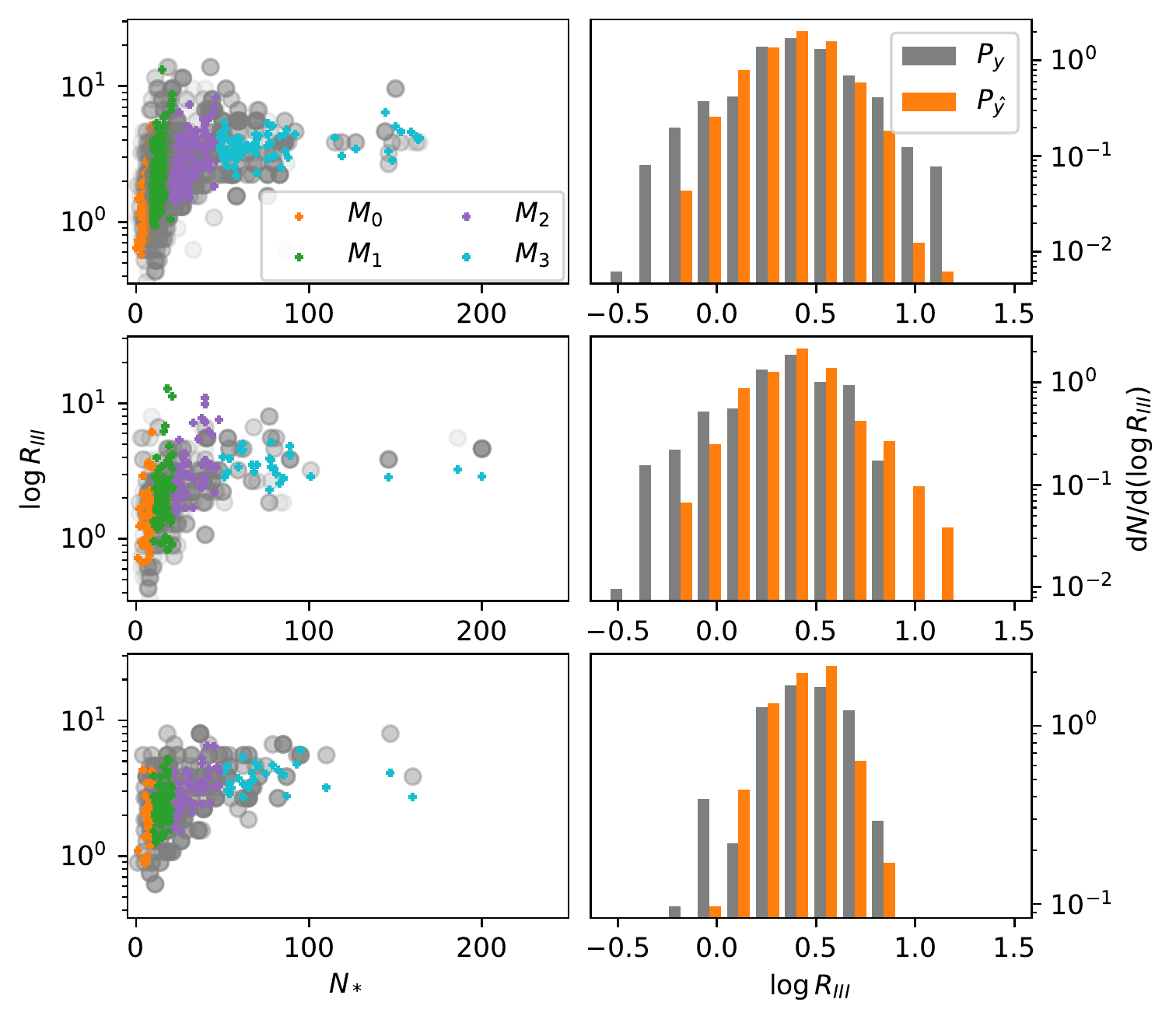}
        \label{fig:19x4_regmodel}}        
        \caption{For predicting the metal radii at two timeframes, 8 and 18 Myr, we present the training (top), validation (middle), and testing (bottom) performance as plots of true radii (grey) with the predictions for each model ($M_0, M_1, M_2, M_3$ in orange, green, purple and cyan, respectively) according to the number of stars in the system ($N_*$). The Left histograms compare the PDF of true radii (grey) with model predictions (orange).}

        \label{fig:regmodels}
    \end{figure*}
    Exemplary results from the 8 Myr and 18 Myr model are shown in Figure \ref{fig:regmodels}; these times are chosen to illustrate the best (8 Myr) and worst (18 Myr) performing models.  On the left side we plot all ground truth points with $\log (R_{III})$ for given $N_*$, overplotted with the model predictions, color coded by which model made the prediction.  While these models score well with $J = 0.189, 0.162$ for 8 and 18 Myr respectively, we can still observe mismatches on the tails of the PDFs for low and high $\log R_{III}$. Note that within the validation and testing dataset, there are no erroneous massive predictions, e.g., predicting $R_{III} > 50$ kpc.  This is largely due to our splitting of the dataset among different models--using a single model for all $N_*$ lead to linear fits that had erroneously high estimates of $R_{III}$ at high $N_*$, however the $M_3$ sub-model, which models high $N_*$, has a nearly flat slope, reducing the errors from very high $N_*$ systems.  As well, a single model struggled to reproduce the low-end of radii from very low $N_*$ systems, making no predictions of $R_{III} < 1$ kpc.  Restricting $M_0$ to predict low-$N_*$ systems allows the fit to have a high slope that can predict both $R_{III} < 1$ kpc and $R_{III} \gtrsim 10$ kpc, while still avoiding falsely large regions that may happen if the same model was responsible for predictions in the higher $N_*$ regime.
    
\begin{figure}
        \centering
        \includegraphics[width=0.46\textwidth]{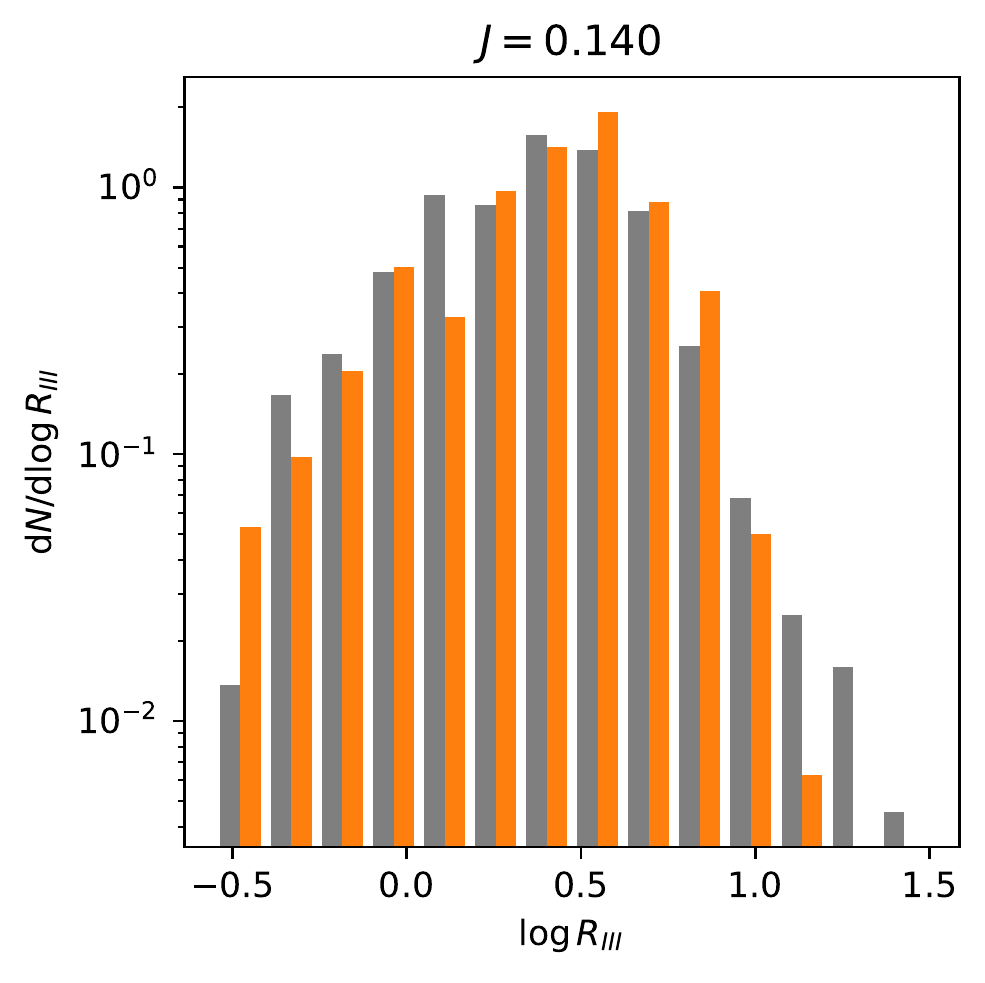}
        \caption{PDF of radii predicted from synthetic dataset (orange) with those from the full ground truth dataset (grey) at $t_{\rm final} = 18$ Myr.  The two PDFs agree well, with $J=0.140$.}
        \label{fig:stresstest}
    \end{figure}

    In the scatter plots, we can identify that the high predictions stem from the $M_1$ and $M_2$ models, however even these results are plausible for the radius of the metal bubble.  As a final stress-test of the model, we generated 3000 samples that match the $N_*$ and mass distribution of the training dataset at 18 Myr, using creation times derived from the SFRs presented in Figure \ref{fig:p3_sfr}.  We compare the PDF of the synthetic sample and full dataset in Figure \ref{fig:stresstest}, where the full dataset is in grey and synthetic dataset in orange.  There is no ground truth for the synthetic dataset (so there are no $R_2, L_1, L_2$ scores), but it produced a PDF that is in agreement with the ground truth dataset, with $J = 0.140$.
    
\section{Conclusions}
    \label{sec:conclusion}
    All halos in the Phoenix suite with $M_{vir} \geq 2\times 10^7$ have \piii star remnants within their radii and have some finite mass of stars (either \piii or \pii).  However, all three simulations also contain several halos without stellar mass above $M_{vir}\geq 10^7~\msun$; these may be candidates for further study to analyze whether they may be super-massive black hole candidates as in the analysis of \cite{regan2017}. At $z\sim 12-14$, all simulations are dominated by \pii star formation, however there continue to be pockets of pristine gas that can form more \piii stars.  As well, we do not observe large-scale enrichment or ionization of the IGM at these redshifts.  Both of these findings are in agreement with prior works \citep{wise2012, xu2016}
    
    \piii stars in the Phoenix suite form in \systems.  These are very diverse, having masses with $M_* \gtrsim 10^3$ $\msun$; 54.7\% of \gii stars are connected to at least one progenitor of each supernovae type.  Inferred by monitoring \piii SFRs at varying radii, it is common to see several \systems in close ($<1.5$ kpc) proximity.  It is also common to see \pii cluster formation near new \piii star formation ($500$ pc$ \lesssim R \lesssim 1500$ pc).  Despite their diversity, a \systems influence is limited to $\lesssim 10$ kpc, with the 95\% CL of maximum radius of influence being $<8$ kpc for both $Z_{III}$ and H$^+$.
    
    Several important observations have been noted regarding \gii stars within the PHX suite: their MDF is not distinguishable from ongoing \pii cluster formation; 78.7\% of \gii clusters were created from gas that had been enriched by all \piii supernova endpoints, 12\% were enriched by only PISNe, and, despite being the endpoint for the IMF peak, only 7.8\% were enriched by only HNe; the highest $Z$ clusters were created from gas enriched by SNe, while the lowest $Z$ gained metals from only PISNe; the number of progenitors is highly variable, averaging only 1 progenitor in PISN case, up to 20 progenitors if all enpoints were found connected to the forming star cluster.  
    
    Both \gii and ongoing \pii stars exhibit low metallicity that includes the ranges of observations of stars found in the JINA database~\citep{abohalima2018}, where most stars have [Fe/H] $\gtrsim -5$.  Although their database has peak metallicity at [Fe/H] $\sim -2.5$, their distribution is inhibited by selection effects at higher metallicity, so is not comparable to this distribution at any metallicity above this level.  The peak of the MDF in the PHX also is similar to the observed metallicity of, e.g., Ursa Minor, with $\langle $[Fe/H]$\rangle = -2.13\pm0.01$ \citep{kirby2013}.   \cite{keller2014} studies the carbon-enhanced metal-poor star SMSS J0313-6708, with [Fe/H] $< -7.1$.  While uncommon, this metallicity represents the extreme low metallicity end of the MDF in Figure \ref{fig:pii_mdf}.  Interestingly, there are no \gii stars with this low metallicity, where the minimum is $Z=-5.92$, but ongoing \pii formation produces several clusters at lower metallicity.
    
    {To evaluate the reasonableness of our \piii SFR and examine the \piii star formation efficiency (SFE), neither of which are constrained observationally, we examine high density cells in comparison to earlier \piii star formation work.  \cite{bromm2002}  presented that the formation of molecular gas clouds proceed at a rate of $\dot M_g = 3.70\times10^{-7}~\msun $ yr$^{-1}$ pc$^{-3}$. $\dot M_g$  is the upper limit on the star formation rate, since a \piii star cannot form without a molecular cloud source.  If we estimate the volume which could form unresolved molecular clouds in PHX256-1 as that which meets the star formation density criteria ($n_b \geq 100$ cm$^{-3}$), 
    % $V \sim 2.26\times10^5$ pc)
    then we find that the PHX256-1 simulation has \piii SFRD $= 7.43\times10^{-9}~\msun$ yr$^{-1}$ pc$^{-3}$ within those regions, implying a \piii SFE of $\sim 2\%$.}  
         
    Data in Table \ref{tab:2g_stats} can be compared to prior works and observational campaigns.  \cite{welsh2020} estimates $5^{+13}_{-3}$ progenitor stars per metal-poor enriched star formation.  Our presentation of $N_{\rm III}$ is largely consistent with their estimate, and highlights the dependence on the IMF to generate reasonable estimates of the number of enriching events. Average star forming halos in \cite{hicks2021} contained $\lesssim$ 20 enriching events, with the most massive halo having $> 100$.  Table \ref{tab:2g_stats} would indicate that the average halo contained a few discrete \systems that led to the final state of the halo, while the most massive may have had several.  \cite{welsh2019} places an upper limit of 70 \piii enriching events in a metal-poor DLA system; our results would indicate that that system would likely have several \systems contained within, or was an exceptionally large single system. Progenitor configurations that lead to \gii star cluster formation are diverse so it would be difficult to analytically build an accurate abundance model for a \gii star.  Future work may be able to track separate metallicity fields from each progenitor type, allowing an estimation of the \gii metal abundances given the composition of progenitors and mixing of the ISM that has occurred.
      
    \system influence can be modelled reasonably well by piecewise linear regression fits, with $R_2\gtrsim 0.4$, reproducing the PDF of observed radii with $J\lesssim 0.2$.  We also find that these models are dependent on the radii to consider from the first stars' formation.  Modelling the radii at longer times requires considering a larger volume centered on the first star--likely due to neighboring formation that affects the original systems ``superbubble".  Along with this consideration, the models struggle with the dynamically active early times of the superbubble, e.g., the first 5 Myr, leaving that these models will be most successfully applied in modelling the influence 8-16 Myr after the \systems' first light.  A more complicated model with parameters for inferring hidden variables or learning from hydrodynamic inputs may be more successful in the short evolution time frame.

\bigskip
\noindent
This research was supported by National Science Foundation CDS\&E grants AST-1615848 and AST-2108076 to M.L.N. The simulations
were performed on the Frontera supercomputer operated
by the Texas Advanced Computing Center (TACC)
with LRAC allocation AST20007. Data analysis was performed on the
Comet and Expanse supercomputers operated for XSEDE by the San Diego
Supercomputer Center.  
 Simulations were performed with Enzo \citep{enzo, Brummel-Smith2019} and analysis with YT \citep{turk2011}, both of which are collaborative open source codes representing efforts from many independent scientists around the world.  The authors thank John Wise for his insight and guidance regarding the \piii and \pii methods in Enzo. 
\bibliographystyle{aasjournal}
\bibliography{bib}
\end{document}